\documentclass[global,twocolumn]{svjourcm}
\usepackage{amsmath}
\usepackage{amsfonts}
\usepackage{amssymb}

\usepackage{graphicx}
\graphicspath{{Figures/}}
\DeclareGraphicsExtensions{.eps,.ps}
\usepackage{epic}

\usepackage{amsmath}
\journalname{}
\begin{document}
  \title{On the Stability and Structural Dynamics of Metal Nanowires}
  \author{ J\'er\^ome B\"urki \and Charles A. Stafford
  }                     
  %
  %
  \institute{Department of Physics, University of Arizona, 1118 E.\ 4th Street,
Tucson, AZ 85721}
  \date{April 22, 2005}
  %
\maketitle

\begin{abstract}
This article presents a brief review of the {\em nanoscale free-electron model},
which provides a continuum description of metal nanostructures.
It is argued that surface and quantum-size effects are the two dominant factors
in the energetics of metal nanowires, and that much of the phenomenology of nanowire stability and structural dynamics can be understood based on the interplay of these
two competing factors.
A linear stability analysis reveals that metal nanocylinders with certain magic conductance values $G=1,3,6,12,17,23,34,42,51,67,78,96,\ldots$ times the conductance quantum are exceptionally stable.
A nonlinear dynamical simulation of nanowire structural evolution reveals a
universal equilibrium shape consisting of a magic cylinder suspended between
unduloidal contacts.
The lifetimes of these metastable structures are also computed.
\end{abstract}

\section{Introduction}\label{intro}

A macroscopic analysis of the mechanical properties of thin metal wires
suggests that it might be difficult to fabricate wires thinner than a few thousand
atoms in cross section:  Consider a cylindrical wire of radius $R$ and
length $L$.  The maximum stress that the wire can sustain before the onset of plastic flow is $\sigma_Y$, the {\em yield strength}.
On the other hand, the surface-induced stress in a thin wire is $\sigma_s/R$, where
$\sigma_s$ is the {\em surface tension}.
If $\sigma_s/R > \sigma_Y$, one would expect the wire to undergo plastic flow and, if $L>2\pi R$, to break up under surface tension, as in the {\em Rayleigh instability} of a column of fluid \cite{Chand81}.
This estimate gives a minimum radius for solidity, $R_{\rm min}= \sigma_s/\sigma_Y$.  The parameters for several simple metals are given in Table \ref{table:nanoscale}.
Plateau realized as early as 1873 that this surface-tension driven instability of a cylinder is unavoidable if cohesion is due solely to classical pairwise interactions
between atoms \cite{Plateau}.

\begin{table}[b]
  \begin{center}
  \begin{tabular}{|cccccc|} \hline 
    Metal & $\sigma_Y$  & $\sigma_s$  & $\gamma_s$
      & $\sigma_s/\sigma_Y$ & $G_{\rm min}$ \\
     & (MPa) & (N/m) &  (pN)  & (nm) & ($G_0$) \\ \hline
     Cu & 210 & 1.5 &  190 &  7.1 & 2300 \\
     Ag & 140 & 1.0 &  154 &  7.4 & 1900 \\
     Au & 100 & 1.3 &  257 &   13 & 5600 \\
     Li & 15 & 0.44 &   99 &   29 & 26000 \\
     Na & 10 & 0.22 &   39 &   22 & 10000 \\
     \hline 
  \end{tabular}
  \end{center}
  \caption{The yield strength $\sigma_Y$ \cite{Metals:reference}, surface tension
    $\sigma_s$ \cite{Tyson77}, and curvature energy $\gamma_s$ \cite{Perdew91} of
    various monovalent metals.
    For a wire of radius $R < \sigma_s/\sigma_Y$, the stress due to surface tension
    exceeds $\sigma_Y$, signalling a breakdown of macroscopic elasticity theory.
    The electrical conductance $G_{\rm min}$ of a ballistic wire of radius
    $R_{\rm min}=\sigma_s/\sigma_Y$ is shown in the rightmost column, in units
    of the conductance quantum $G_0=2e^2/h$.
    Note that $G/G_0$ is approximately equal to the number of atoms that fit within
    the cross section for monovalent metals.
    \label{table:nanoscale}}
\end{table}

A great deal of experimental evidence has accumulated over the past decade, however, indicating that metal wires considerably thinner than the above estimate
can be fabricated by a number of different techniques
\cite{Rubio96,Untiedt97,Kondo97,Ohnishi98,Yanson98,Yanson99,Kondo00,Rodrigues00,Yanson00,Yanson01,Yanson01a,Rodrigues02b,Oshima03,Oshima03a,Diaz03}.
Even wires with lengths significantly exceeding their circumference were found to be remarkably stable \cite{Kondo97,Ohnishi98,Yanson98,Rodrigues02b,Oshima03,Oshima03a},
indicating that some new mechanism must intervene to prevent their breakup.

A clue to the resolution of this problem was provided by the observation
of electron-shell structure in conductance histograms of
alkali metal point contacts \cite{Yanson99,Yanson00,Yanson01,Yanson01a}.
Like the surface tension, quantum-size effects arising from the confinement
of the conduction electrons within the cross-section of the wire become increasingly important as the wire is scaled down to atomic dimensions.
In fact, a linear stability analysis \cite{Kassubek01,Zhang03}
of ultrathin metal wires within the free-electron model found that the Rayleigh instability can be completely suppressed in the vicinity of certain {\em magic radii}.

In this article, we argue that surface and quantum-size effects are the two dominant factors in the energetics of {\em metal nanowires}, that is, metal wires with
$R < R_{\rm min}$.
We show that much of the phenomenology of nanowire stability and structural dynamics can be understood based on the interplay of these two competing factors.

This article is organized as follows:  In Sec.\ \ref{sec:model}, we describe our continuum structural model for metal nanowires.
A linear stability analysis of metal nanowires is presented in  Sec.\ \ref{sec:stability}.
Section \ref{sec:dynamics} describes the structural evolution of a metal nanowire from a random initial configuration to a universal equilibrium shape.
The thermally-activated decay of metal nanowires is discussed in Sec.\ \ref{sec:lifetime}.
Some concluding remarks are given in Sec.\ \ref{sec:conclusions}.

\section{The Nanoscale Free-Electron Model}\label{sec:model}

Guided by the importance of conduction electrons in the cohesion of metals,
and by the success of the jellium model in describing metal clusters \cite{Brack93},
the nanoscale free-electron model (NFEM) \cite{Stafford97a} replaces the metal ions
by a uniform, positively charged background that provides a confinement potential for the electrons.
The electron motion is free along the wire, and confined in the transverse directions.
Due to the excellent screening \cite{Kassubek99,Zhang05} in metal wires with $G>G_0$,
electron-electron interactions can in most cases be neglected.
The surface properties of various metals can be fit by using appropriate surface boundary conditions \cite{Garcia-martin96,Urban04}.

The NFEM is especially suitable for alkali metals, but is also adequate to describe shell effects due to the conduction-band $s$-electrons in other monovalent metals, such as gold.
The experimental observation of a crossover from atomic-shell to electron-shell
effects with decreasing radius in both metal clusters \cite{Martin96} and nanowires \cite{Yanson01,Yanson01a} justifies {\it a posteriori} the use of the NFEM in the later regime.

A nanowire connecting two macroscopic electrodes is an open quantum system, for which the Schr\"odinger equation is most naturally formulated as a scattering problem.
Transport properties can be obtained from the scattering matrix using Landauer-type formulas \cite{Stafford97a,Burki99,Burki99a}, while cohesive properties require the computation of the grand canonical potential of the electrons.
The later can also be expressed in terms of the scattering matrix \cite{Stafford97a}, or
calculated semiclassically \cite{Stafford99} in terms of geometrical quantities and a sum over classical periodic orbits, as presented in Sect.\ \ref{sec:Omega_e}.

Motivated by the argument presented in Table \ref{table:nanoscale},
the ionic degrees of freedom in the wire are modeled as an incompressible, irrotational fluid \cite{Zhang03,Burki03}.
In the Born-Oppenheimer approximation, the electronic free energy serves as the potential energy for the ions.
The ionic dynamics may then be modeled via a surface self-diffusion equation \cite{Burki03}, as presented in Sec.\ \ref{sec:diffusion} or, taking thermal fluctuations into account, via a classical Ginzburg-Landau stochastic field theory \cite{Burki04b}, as presented in Sec.\ \ref{sec:fluct}.

\subsection{Electronic Energy Functional}\label{sec:Omega_e}

\begin{figure}[b]
   \centering
   \includegraphics[width=8.5cm,angle=0]{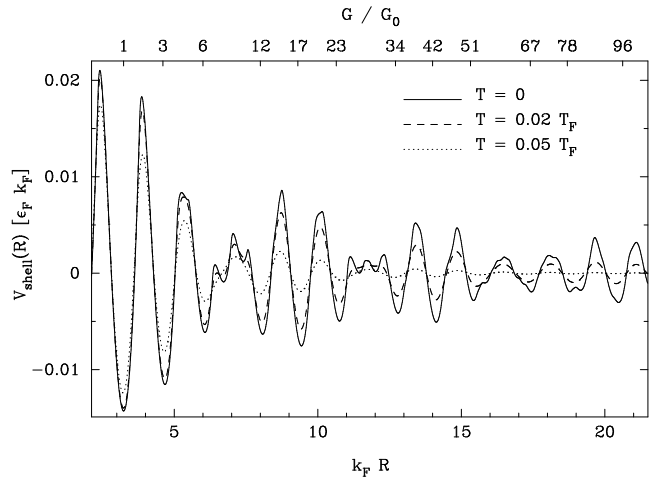}
   \caption{Electron-shell potential $V_{shell}(R,T)$ at zero and two finite
     temperatures, which correspond respectively to 1000K and 2500K for Na.
     The electrical conductance values of the magic cylindrical wires
     are indicated on the upper axis.
     \label{fig:potential}}
\end{figure}

Restricting ourselves to axisymmetric structures, the grand canonical potential for the electrons $\Omega_e$ becomes a functional of the radius $R(z)$ of the wire.
Using the Weyl expansion \cite{Brack97}, $\Omega_e$ can be expressed in terms of geometrical quantities such as the volume ${\cal V}$, surface area ${\cal S}$,
and integrated mean curvature ${\cal C}$ of the wire's surface, plus an electron-shell correction,
\begin{equation}\label{eq:omega}  
  \Omega_e\bigl[R(z),T\bigr] = -\omega{\cal V} + \sigma_s{\cal S} - \gamma_s{\cal C}
    + \int_0^L\!\!\text{d}z\,V_{shell},
\end{equation}  
where $-\omega$ is the bulk value per unit volume, $\sigma_s$ is the surface tension, $\gamma_s$ is a curvature-energy density, and $V_{shell}\bigl(R(z),T\bigr)$ is a mesoscopic electron-shell potential, shown in Fig.\ \ref{fig:potential}.
The parameters $\sigma_s$ and $\gamma_s$, tabulated for various metals in Table \ref{table:nanoscale}, depend on the details of the interaction-dependent surface confinement potential \cite{Garcia-martin96,Urban04,Stafford99},
but can be taken as phenomenological material-dependent parameters (along with $\omega$) in our model.
The leading-order electron-shell correction is, however, independent of the Coulomb interaction \cite{Kassubek99,Zhang05,Stafford99},
and therefore insensitive to the details of the confinement potential.

The geometrical quantities ${\cal S}=\int_0^L\!\!\text{d}z\,\partial{\cal S}$ and ${\cal C}=\int_0^L\!\!\text{d}z\,\partial{\cal C}$ are given by
\begin{equation}\label{eq:surf}
  \partial{\cal S}\bigl[R(z)\bigr]=2\pi R(z)\sqrt{1+(\partial_zR)^2},
\end{equation}
and
\begin{equation}\label{eq:curv}
  \partial{\cal C}\bigl[R(z)\bigr]=\pi\left(1
	- \frac{R \, \partial^2_{z}R}{1+(\partial_zR)^2}\right),
\end{equation}
where $\partial_z=\partial /\partial z$.

Approximating the confining potential by a hard wall at the surface of the wire,
the electron-shell potential $V_{shell}$ can be expressed in terms of a Gutzwiller-type
trace formula \cite{Burki03}
\begin{equation}\label{eq:gutzwiller}  
  V_{shell}(R,T) = \frac{2\varepsilon_F}{\pi}
    \sum_{w=1}^{\infty}\sum_{v=2w}^{\infty} a_{vw}(T)
    \frac{f_{vw}\cos\theta_{vw}}{v^2L_{vw}},
\end{equation}  
where the sum includes all classical periodic orbits $(v,w)$ in a disk
billiard \cite{Brack97}, characterized by their number of vertices $v$
and winding number $w$, $L_{vw}=2vR\sin(\pi w/v)$ is the length of
orbit $(v,w)$, and $\theta_{vw}=k_FL_{vw}-3v\pi/2$.  The factor
$f_{vw}=1\text{ for } v=2w, 2$ otherwise, accounts for the invariance
under time-reversal symmetry of some orbits, and $a_{vw}(T) =
\tau_{vw}/\sinh{\tau_{vw}}$ ($\tau_{vw}=\pi k_FL_{vw}T/2T_F$) is a
temperature-dependent damping factor.

$V_{shell}(R)$ exhibits deep minima as a function of $R$ (see Fig.\ \ref{fig:potential}), suggesting that some radii are strongly favored, which is confirmed by the stability analysis of Sec.\ \ref{sec:stability}.
Note that room temperature is small compared to the Fermi temperature $T_F=\varepsilon_F/k_B$, (in particular, $T/T_F=0.008$ at
$T=300\mbox{K}$ for Na), so that the finite-temperature electron-shell potential is essentially indistinguishable from its zero-temperature limit at
experimental temperatures.

\subsection{Ionic Energetics}\label{sec:Omega_a}

In the Born-Oppenheimer approximation, the electronic energy\  (\ref{eq:omega}) acts as a potential energy for the ionic background.
The wire can exchange atoms with the macroscopic contacts via surface self-diffusion, so the grand canonical ensemble has to be used for the ionic background as well, leading to an ionic grand canonical potential
\begin{equation}\label{eq:energy_atoms}
  \Omega_a = \Omega_e - \mu_a {\cal N}_a,
\end{equation}
where ${\cal N}_a={\cal V}/{\cal V}_a$ is the number of positive ions in the wire (${\cal V}_a=3\pi^2/k_F^3$ is the volume of an atom), and $\mu_a$ is the chemical potential for a surface atom in the wire.
Using Eqs.\ (\ref{eq:omega}--\ref{eq:curv}), the ionic free energy (\ref{eq:energy_atoms}) becomes
\begin{multline}\label{eq:omega2}
  \Omega_a = \int\text{d}z\Bigg[2\pi\sigma_s
           R(z,t)\sqrt{1+(\partial_zR)^2}  \\
	     -\pi\gamma_s + V_{shell}(R,T) \Bigg]
	     -(\omega + \mu_a/{\cal V}_a){\cal V},
\end{multline}
where only the leading-order term in the curvature energy is included.
The chemical potential $\mu_a$ is obtained by calculating the change in the
energy (\ref{eq:omega}) with the addition of an atom at point $z_0$,
$\mu_a(z_0) = \Omega_e\bigl[R(z)+c\delta(z-z_0),T\bigr] - \Omega_e\bigl[R(z),T\bigr]$, where $c={\cal V}_a/2\pi R(z)$ is chosen so that the volume of an atom is added:
\begin{multline}\label{eq:mu}  
  \mu_a(z) = -\omega {\cal V}_a
    + \frac{{\cal V}_a}{2\pi R}\left(
      \frac{2\sigma_s\partial{\cal C}[R(z)]}{\sqrt{1+(\partial_z R)^2}}
      + \frac{\partial V_{shell}}{\partial R}\right).
\end{multline}  

\subsection{Structural Dynamics}
\subsubsection{Surface self-diffusion}\label{sec:diffusion}

Since a large fraction of the atoms in a nanowire are on the surface, surface self-diffusion is the dominant mechanism of ionic motion \cite{Burki03}.
The dynamics derive from ionic mass conservation:
\begin{equation}\label{eq:diffusion}  
  \frac{\pi}{{\cal V}_a}\frac{\partial R^2(z,t)}{\partial t}
	+ \frac{\partial}{\partial z}\bigl[2\pi R(z,t)J_z(z,t)\bigr] = 0,
\end{equation}  
where the $z$-component of the surface current density is given by Fick's law:
\begin{equation}\label{eq:current} 
  J_z = -\frac{\rho_SD_S}{k_B T}\frac{1}{\sqrt{1+(\partial_zR)^2}}
        \frac{\partial\mu_a}{\partial z}.
\end{equation}  
Here $\rho_S$ and $D_S$ are the surface density of ions and
the surface self-diffusion coefficient, respectively.
The precise value of $D_S$ for most metals is not known, but it can be removed from the evolution equation by rescaling time to the dimensionless variable $\tau=(\rho_SD_ST_F/T)t$.
For comparison to experimental time scales, one can estimate that for quasi-one-dimensional diffusion $D_s \approx \nu_D a^2 \exp(-E_s/k_B T)$,
where $\nu_D$ is the Debye frequency, $a$ is the lattice spacing, and $E_s$ is an
activation energy comparable to the energy of a single bond in the solid.
Our non-linear dynamical model, Eqs.\ (\ref{eq:mu}--\ref{eq:current}),
differs from previous studies of axisymmetric surface self-diffusion
\cite{Coleman95,Eggers98,Bernoff98} by the inclusion of electron-shell effects [last term of Eq.\ (\ref{eq:mu})], which fundamentally alter the dynamics.

\subsubsection{Thermal fluctuations}\label{sec:fluct}

The diffusive dynamics of the previous subsection describe relaxation toward structures of lower free energy.
Once an equilibrium configuration (i.e., a local minimum of the free energy) is attained, however, fluctuations about this configuration will dominate the dynamics, limiting the dwell time of the system in this local minimum.
As shown in Sec.\ \ref{sec:dynamics}, the equilibrium configurations consist of stable cylindrical nanowires in diffusive equilibrium with unduloid-like contacts \cite{Burki03,Burki04a}.
We therefore study fluctuations of the form
\begin{equation}\label{eq:radius}
  R(z,t) \equiv R_0 + \phi(z,t),
\end{equation}
where $R_0$ is the radius of a stable cylinder of length $L$.

The energy (\ref{eq:omega2}) can be expanded as a series in $\phi$.
For the magic cylinders, corresponding to minima of $V_{shell}(R_0)$ (c.f.\ Fig.~\ref{fig:potential}), the chemical potential for the exchange of atoms between the wire and the contacts reduces to
\begin{equation}\label{eq:mu_a_stable}
  \frac{\mu_a}{{\cal V}_a} = \frac{\sigma_s}{R_0} - \omega.
\end{equation}
Keeping only the leading-order terms in $\partial_z \phi$, one gets
$\Omega_a = \Omega_a(R_0) + {\cal H}[\phi]$, where $\Omega_a(R_0)$ is the
energy of an unperturbed cylinder of radius $R_0$ and
\begin{equation}\label{eq:energy}
  {\cal H}[\phi] = \int_0^L\!\!\text{d}z\left[
  	\frac{\kappa}{2}(\partial_z\phi)^2 + V(\phi)\right].
\end{equation}
Here $\kappa=2\pi\sigma_s R_0$ and
\begin{equation}\label{eq:Vtilde}
  V(\phi) \equiv V_{shell}(R_0+\phi) - V_{shell}(R_0)-\frac{\pi\sigma_s}{R_0}\phi^2.
\end{equation}

The problem of stability of nanowires against thermal fluctuations can be studied as a one-dimensional Ginzburg--Landau scalar field theory,
perturbed by weak spatiotemporal noise, in a domain of finite extent (see \cite{Burki04b} and references therein):
The fluctuations of the nanowire radius $\phi$ are treated as a classical field on a
one-dimensional spatial domain $[0,L]$.  Its dynamics are governed by the
stochastic Ginzburg--Landau equation
\begin{equation}\label{eq:GL}
  \frac{\partial \phi(z,t)}{\partial t} =
      \kappa \frac{\partial^2 \phi}{\partial z^2}
      - \frac{\partial V}{\partial\phi} + (2T)^{1/2}\xi(z,t),
\end{equation}
where $\xi(z,t)$ is unit-strength spatiotemporal white noise, satisfying
$\langle\xi(z_1,t_1)\xi(z_2,t_2)\rangle=\delta(z_1-z_2)\delta(t_1-t_2)$.
In Eq.\ (\ref{eq:GL}), time is measured in units of a microscopic
timescale describing the short-wavelength cutoff of the surface
dynamics \cite{Zhang03}, which is given to within a factor of order
unity by the inverse Debye frequency $\nu_D^{-1}$.
The zero-noise dynamics is ``gradient,''  that is,
at zero temperature $\dot\phi=-\delta{\cal H}/\delta\phi$,
where ${\cal H}[\phi]$ is given by Eq.\ (\ref{eq:energy}).
Eq.\ (\ref{eq:GL}) represents a considerable simplification compared to the volume-conserving dynamics of Eq.\ (\ref{eq:diffusion})
(which involves derivatives up to $\partial^4_z R$),
and makes possible an analytical treatment of thermal fluctuations.

\section{Linear Stability of Cylinders}\label{sec:stability}

The linear stability of a structure is determined by studying the change of energy induced by a small perturbation:
If any one perturbation decreases the energy, the structure is unstable, while it is stable if all perturbations increase the energy.

The most general perturbation of a cylinder of radius $R_0$ and length $L$ is
\begin{equation}\label{eq:rm_z}
  R(z,\phi) = R_0 + \lambda\sum_{m}\sum_q b_m(q)e^{i(qz+m\phi)},
\end{equation}
where $b_m(q)=b_{-m}(-q)^*$.  For simplicity, we impose periodic boundary conditions, so that $q$ is an integer multiple of $2\pi/L$.
Since the total number of atoms in the system is unchanged by the perturbation, $b_0(0)$ is related to the other coefficients by volume conservation
\begin{equation}
  b_0(0) = -\frac\lambda{R_0}\sum_m \sum_{q > 0} |b_m(q)|^2 + {\cal O}(\lambda^2),
\end{equation}
and may be eliminated.  Other constraints \cite{Urban04} may be utilized to account for confinement potentials more general \cite{Garcia-martin96}
than the hard walls considered in the present article, but do not lead to a qualitative change in the stability analysis.

The energy change (per unit length) under such a perturbation is found to be
\begin{equation}\label{eq:dOmega}
  \frac{\Delta\Omega_e}{L} = \lambda^2 \sum_m \sum_{q>0}
    \alpha_m(q;R_0,T)|b_m(q)|^2 + {\cal O}(\lambda^3),
\end{equation}
where the mode stiffness $\alpha_m(q)$ is given by
\begin{eqnarray}\label{eq:stiffness}
  \alpha_m(q;R,T) & = & (m^2-1)\frac{2\pi\sigma_s}{R}
  	+ 2\pi(\sigma_s R-\gamma_s)q^2 \nonumber \\
        & & \mbox{} + \delta\alpha_m(q;R,T),
\end{eqnarray}
and $\delta\alpha_m$ is a mesoscopic electron-shell correction. 

Neglecting for the moment the mesoscopic correction $\delta\alpha_m(q)$, we find that the perturbation can only lead to an instability for
$m=0$ and $qR_0 < \big(1-\gamma_s/\sigma_s R_0\big)^{-1/2}\approx 1$, which is the criterion for the classical Rayleigh instability
\cite{Chand81}.
Note that $\sigma_s R_0 > \gamma_s$ for all physically meaningful radii (c.f.\ Table \ref{table:nanoscale}).
Any perturbation breaking axial symmetry is classically unfavorable, and we will therefore consider only axisymmetric perturbations ($m=0$) in the rest of this paper.

\begin{figure}[t]
   \centering
   \includegraphics[width=0.91\columnwidth]{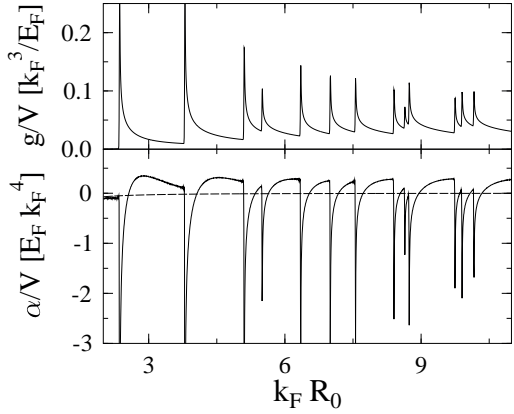}
   \caption{Density of states at the Fermi energy $g$ (top)
   	and mode-stiffness for axisymmetric deformations $\alpha\equiv\alpha_0(q)$
	(bottom), normalized by the volume ${\cal V}$ of the wire.
	The perturbation wavevector is given by $qR_0=1$, so that the surface
        contribution (dashed curve) to $\alpha$ is nearly zero.
   \label{fig:stiffness}}
\end{figure}

Using semiclassical perturbation theory, the electron-shell correction to the mode stiffness for axisymmetric deformations was found to be independent of $q$ \cite{Kassubek01,Zhang03},
\begin{equation}\label{eq:meso_stiffness}
  \delta\alpha_0(R,T) = \left(\frac{\partial^2}{\partial R^2}
  	- \frac1R\frac{\partial}{\partial R}\right)V_{shell}(R,T).
\end{equation}
This turns out to be true only in the semiclassical approximation:
A fully quantum-mechanical stability analysis \cite{Urban03} reveals that long wires undergo a Peierls-type instability at $q=2k_F^{(\nu)}$, where $k_F^{(\nu)}$ is the Fermi wavevector for subband $\nu$.
However, the semiclassical results are found to provide a good approximation as long as the temperature is not too low, and the wires are not too long \cite{Urban03}.
The total mode stiffness $\alpha_0(q=1/R_0)$ in the semiclassical approximation
is shown in Fig.~\ref{fig:stiffness}, together with the density of states $g(\varepsilon_F)$.
The perturbation wavevector was chosen so that the surface contribution to $\alpha_0$ (dashed curve) is nearly zero.
Fig.\ \ref{fig:stiffness} shows that near the thresholds to open new conducting channels, where the density of states is large, the wire is very unstable ($\alpha_0 <0$).  However, in between the subband thresholds, the shell correction {\em stabilizes} the wire ($\alpha_0 >0$).

\begin{figure}[t]
   \centering
   \includegraphics[width=0.99\columnwidth]{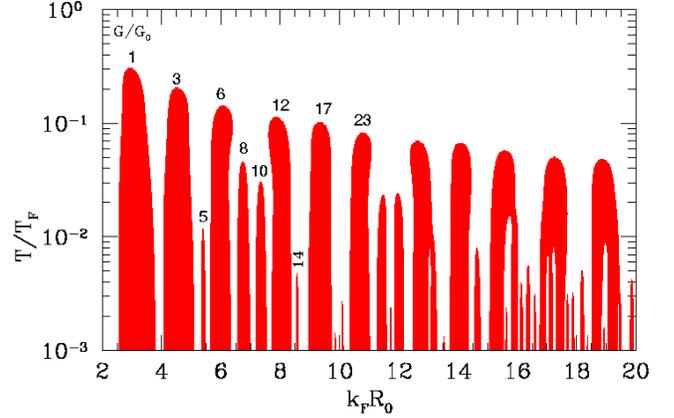}
   \caption{Stability diagram for cylindrical metal nanowires.
     Dark areas indicate stability with respect to small perturbations,
     $A(R_0,T) > 0$.
     The quantized conductance values of some of the stable wires are indicated.
     \label{fig:stability}}
\end{figure}

According to Eqs.\ (\ref{eq:stiffness}) and (\ref{eq:meso_stiffness}), the most unstable mode is $m=0$, $q=0$.
The stability of the wire is thus determined by the sign of the stability coefficient
$A(R_0,T)  \equiv \alpha_0(q=0;R_0,T)$,
\begin{equation}\label{eq:stab_coeff}
  A(R,T) = -\frac{2\pi \sigma_s}{R} + \left(\frac{\partial^2}{\partial R^2}
    - \frac1R\frac{\partial}{\partial R}\right)V_{shell}(R,T).
\end{equation}
For $A > 0$, the wire is stable with respect to all small perturbations, while
the wire is unstable for $A < 0$.
The stability diagram so determined is shown in Fig.~\ref{fig:stability}.
Competition between surface tension and electron-shell effects leads to a complex landscape of stable fingers and arches extending up to very high
temperatures: cylindrical wires whose electrical conductance is a magic number 1, 3, 6, 12, 17, 23,... times the conductance quantum are predicted to be stable with respect to small perturbations up to temperatures well above the bulk melting temperature
$T_M \approx 0.01 T_F$.  This finding suggests that metal nanowires are remarkably robust structures.
Indeed, the principal stable zones shown in Fig.\ \ref{fig:stability} were found \cite{Zhang05} to persist up to bias voltages $eV \geq 0.1 \varepsilon_F$, implying that these wires can support current densities greater than $10^{10}\mbox{A}/\mbox{cm}^2$, which would vaporize a macroscopic wire.
(See Sec.\ \ref{sec:lifetime} for a discussion of nanowire lifetime as a function of temperature).
In Fig.\ \ref{fig:stability}, the values \cite{Stafford99}
$\sigma_s = \varepsilon_F k_F^2/80\pi$ and $\gamma_s=4\varepsilon_F k_F/45\pi^2$, appropriate for alkali metals \cite{Zhang03}, were used.
For larger values of $\sigma_s$ (e.g., for noble metals), the maximum temperatures (in units of $T_F$) of the stable fingers are reduced somewhat, but the stability diagram is qualitatively the same.

The fact, illustrated in Fig.\ \ref{fig:stability}, that electron-shell effects can overcome the surface-tension driven instability of a cylinder is rather remarkable.  The surface contribution to $\Omega_e$ is ${\cal O}(k_F R_0)$, while the electron-shell contribution (\ref{eq:gutzwiller}) is ${\cal O}(k_F R_0)^{-1}$.
For a typical radius $k_F R_0 =10$, the shell-correction to the energy is thus
two orders of magnitude smaller than the surface energy!  Stability is not determined by the energy directly, however, but rather by the {\em convexity} (or lack thereof) of the energy functional, which involves the second derivative with respect to $R_0$
[c.f.\ Eq.\ (\ref{eq:stab_coeff})].  Because $V_{shell}$ is a rapidly oscillating function of $R_0$, its second derivative actually has the same characteristic size as the surface contribution to the stability coefficient [first term on the r.h.s.\ of
Eq.\ (\ref{eq:stab_coeff})].

Cylinders are special in this respect, because the term ${\cal O}(\lambda)$ in Eq.\ (\ref{eq:dOmega}) vanishes exactly by symmetry.
For a more general shape (such as a wire with an elliptical cross section \cite{Urban04}) to be stable, the first variations of the surface energy and electron-shell energy, which do not have the same characteristic size, must cancel.  This is only possible for small $k_F R_0$ and/or for small deviations from cylindrical symmetry.  Thus cylinders represent about 75\% of the experimentally observable
\cite{Urban04b} (most stable) wires, while structures of lesser symmetry represent only about 25\% of the total.

Further insight into the stability criterion $A(R_0,T)>0$ is provided by the identity
\begin{equation}\label{eq:identity}
  \left.\frac{\partial \mu_{\rm cyl}(R_0,T)}{\partial R_0}\right|_T
     = \frac{{\cal V}_a}{2\pi R_0} A(R_0,T),
\end{equation}
where $\mu_{\rm cyl}(R_0,T)$ is given by Eq.\ (\ref{eq:mu}) with $R(z)=R_0$.
The wire can lower its free energy via phase separation into thicker and thinner segments if and only if $\partial \mu_{\rm cyl}/\partial R_0 < 0$.
$A < 0$ therefore corresponds to an inhomogeneous phase \cite{Zhang03}, while $A>0$ corresponds to a homogeneous phase.
This is confirmed by dynamical simulation of weakly perturbed stable and unstable cylinders, the later evolving into an inhomogeneous wire \cite{Burki03}.

\section{Evolution toward Equilibrium}\label{sec:dynamics}

\begin{figure}[t]
   \centering
   \includegraphics[width=0.99\columnwidth]{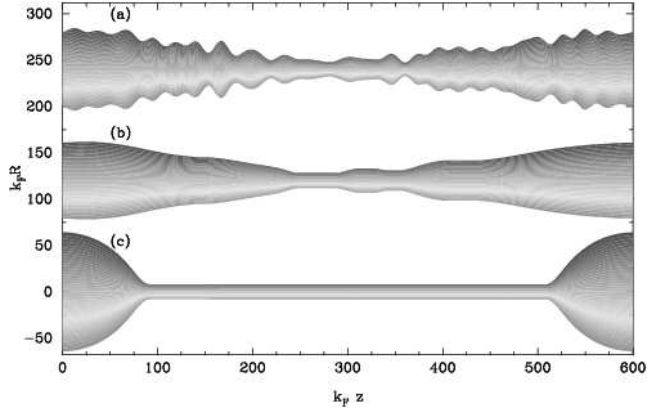}
   \caption{Equilibration of an initially random nanowire:
      (a) initial shape; (b) $\tau=2\times10^4$;
      (c) $\tau=3\times10^7$, equilibrium structure with $G=12\,G_0$.
   \label{fig:Equilibrium}}
\end{figure}

In this section, we use the diffusion equation (\ref{eq:diffusion}) to study the equilibrium shapes of metal nanowires, as well as the approach to equilibrium.
Figure~\ref{fig:Equilibrium} shows three stages of the typical evolution \cite{Burki03,Burki04a} of an initially random (a) nanowire:
After a relatively short time (b), the short-wavelength surface roughness is smoothed out, leaving a few cylindrical segments, connected by kinks;
Eventually, all kinks propagate outward and coallesce,
yielding an equilibrium shape (c) consisting of a cylindrical wire suspended between two thicker contacts.

\begin{figure}[t]
   \setlength\unitlength{\columnwidth}
   \begin{picture}(1,1.1)
    \put(0,0){\includegraphics[width=0.99\columnwidth]{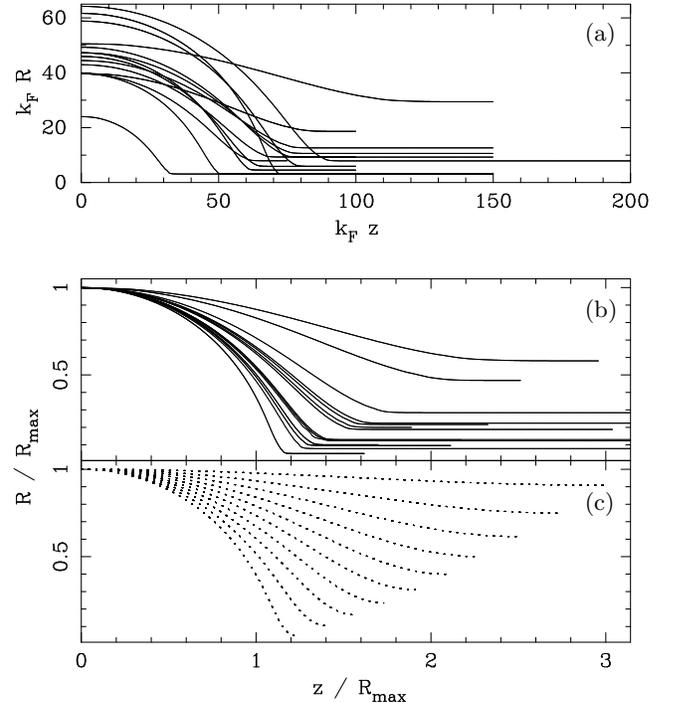}}
    \put(0.89,1.03){(a)}
    \put(0.89,0.6){(b)}
    \put(0.89,0.3){(c)}
   \end{picture}
   \caption{(a) Radius $R(z)$ of the equilibrium shapes for 14 simulations
   	starting from random initial shapes; the equilibrium shapes being symmetric
	(although the initial shapes are not), we only show $R(z)$
	for $z\in[0,L/2]$.
	(b) Same shapes rescaled by their maximum radius $R_{max}$.
	(c) Series of Delaunay unduloids of various curvature.
   \label{fig:Universality}}
\end{figure}

Several such simulations starting from various initial configurations, with conductance ranging  from $1$ to $200\,G_0$, and lengths $200 \leqslant k_FL \leqslant 600$, all evolved to equilibrium structures consisting of one of the stable cylinders found in Sec.\ \ref{sec:stability}, connecting two quasi-spherical contacts (see Fig.~\ref{fig:Universality}(a)).
The shape of the contact is actually a close approximation to a {\em Delaunay unduloid of revolution} \cite{Bernoff98}, which is a surface of constant mean curvature, and is an unstable steady state of diffusion equation (\ref{eq:diffusion})
without the shell-effect term.
This is illustrated in Fig.~\ref{fig:Universality}, comparing the equilibrium wires, rescaled (b) by their maximum radius $R_{max}$ to a series of unduloids (c) of various mean curvature.
The curvature of the unduloid is determined solely by the ratio of the radius of the cylindrical part to $R_{max}$,
and not by the conductance of the wire, or its length.
In our case, the deep minima of the electron-shell potential, Fig.~\ref{fig:potential}, pin the unduloid at its connection with the cylindrical part, thus stabilizing it.
In fact, if one switches off the electron-shell potential in the simulations, the equilibrated wires break apart, as expected from the Rayleigh instability.
The breaking is found to happen first at the junction between the cylinder and the lead,
suggesting that it is the weak point of the equilibrium structure.

This suggests that the natural evolution of a nanowire, at a temperature sufficient for surface atoms to diffuse, is to form a cylinder, thus providing an explanation of the observation of long, almost perfect cylindrical Au nanowires in transmission electron microscope (TEM) experiments \cite{Kondo97,Kondo00,Rodrigues00,Oshima03}.
The same type of simulation can be used to understand the thinning process observed in TEM experiments \cite{Oshima03a}, where the wire diameter is seen to decrease step by step through the propagation of kinks along the wire.

\section{Lifetimes of Metastable Cylinders}\label{sec:lifetime}

The equilibrium nanowire structures determined in the preceding sections are stable with respect to small perturbations, and represent local minima of the free energy functional (\ref{eq:omega2}).
However, large perturbations induced e.g.\ by thermal fluctuations can drive the nanowire out of such a minimum, leading to a finite lifetime of these metastable structures.  In this section, we use the stochastic model \cite{Burki04b} derived in Sec.\ \ref{sec:fluct} to study this process.

The statistical properties of the stochastically evolving field~$\phi$, Eq.\ (\ref{eq:radius}), are described by equilibrium statistical mechanics.
At nonzero temperature, thermal fluctuations can induce transitions between stable states (i.e., local minima) of the potential $V(\phi)$, Eq.\ (\ref{eq:Vtilde}).
Such transitions occur via nucleation of a ``droplet'' of one stable configuration
in the background of the other, subsequently quickly spreading to fill the
entire spatial domain.  When the noise is weak, i.e., at low temperatures
(compared to the barrier height) most fluctuations will not succeed in
nucleating a new phase; it is far more likely for a small droplet to shrink
and vanish.

A transition state must go ``uphill'' in energy from each stable field configuration.  Because of exponential suppression of fluctuations as their energy increases, there is
at low temperature a preferred transition configuration (saddle) that lies between adjacent minima.  These are the nucleation pathways.
By time-reversal invariance, they are time-reversed zero-noise ``downhill'' trajectories~\cite{MS93}.
At low temperatures, the expected waiting time of the order parameter $\phi$ in a
basin of attraction is an exponential random variable, as is typical of slow rate processes.
The activation rate is given in the $T\to0$ limit by the Kramers formula
\begin{equation}\label{eq:Kramers}
  \Gamma\sim \Gamma_0\exp(-\Delta E/T)\, .
\end{equation}
Here the activation barrier $\Delta E$ is the energy of the transition state minus that of the stable state, and $\Gamma_0$ is the rate prefactor.
The quantities $\Delta E$ and $\Gamma_0$ depend on the details of the
potential, on the length~$L$, and on the choice of boundary conditions at
the endpoints $z=0$ and~$z=L$.
Based on the equilibrium structures found in Sec.\ \ref{sec:dynamics}, we employ Neumann boundary conditions, $0=\partial_z \phi(z,t)|_{z=0,L}$.
These boundary conditions force nucleation to begin, preferentially, at the endpoints, consistent with experimental observations \cite{Oshima03a}.

\begin{figure}[t]
\begin{center}
  \includegraphics[width=0.95\columnwidth]{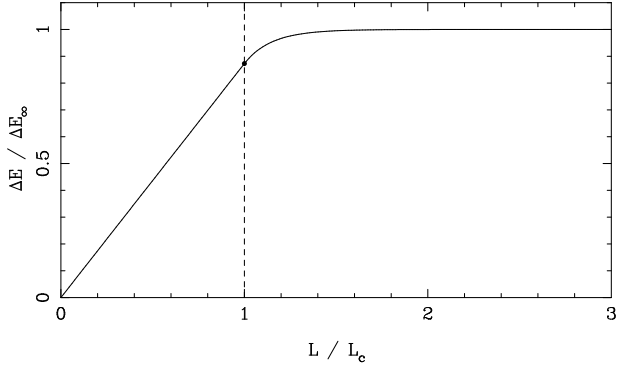}
\end{center}
  \caption{Escape barrier $\Delta E$ as a function of the wire length $L$.
        Here $L_c$ is the critical length at which the transition state bifurcates.
	\label{fig:lifetime}}
\end{figure}

Equation (\ref{eq:GL}) with the potential (\ref{eq:Vtilde}) can not in general be solved analytically, but most minima of the potential $V(\phi)$ can be locally approximated by a cubic potential
\begin{equation}\label{eq:cubic}
  V^{(\pm)}(\phi) = -\alpha \tilde\phi_\pm + \frac{\beta}{3} \tilde\phi^3_\pm\, ,
\end{equation}
where $\tilde\phi_\pm=\sqrt{\alpha/\beta}\mp \phi$ ($\alpha, \beta > 0$).
The potential $V^{(-)}$ ($V^{(+)}$) biases fluctuations toward smaller (larger) radii.

Fig.~\ref{fig:lifetime} shows the escape barrier $\Delta E$ 
as a function of the wire length \cite{Burki04b}: Below a critical length~$L_c$, the transition state is a spatially constant field configuration, and the escape barrier grows linearly with the wire length $L$.  However, at~$L=L_c$ it
bifurcates into a spatially varying {\em instanton} configuration with characteristic size $\sim L_c$, so that $\Delta E$ becomes length-independent for $L\gg L_c$.

Our continuum dynamical model thus predicts that the lifetime $\tau$ of a metastable
cylindrical nanowire of length greater than the critical length $L_c$ saturates with an
escape barrier given by $\Delta E_\infty = \lim_{L\rightarrow \infty} \Delta E$.
In terms of the physical parameters defining the cubic potential\ (\ref{eq:cubic}), the critical length $L_c = \frac{\pi}{\sqrt{2}}\kappa^{1/2}/(\alpha\beta)^{1/4}$ and
$\Delta E_\infty=\frac{12\sqrt{2}}{5} \kappa^{1/2} \alpha^{5/4}/\beta^{3/4}$.
The lifetimes $\tau=1/\Gamma$ for several cylindrical sodium nanowires, calculated using the best cubic-polynomial fits to the potential (\ref{eq:Vtilde}), are tabulated in Table \ref{tab:lifetime}.
Note that for a wire with $G/G_0 > 1$, the lifetime $\tau$ may not be the typical time before the wire breaks, but rather a switching time between two different metastable wires with different conductance values.

\begin{table}[t]
  \caption{The lifetime $\tau$ (in seconds) for various cylindrical sodium
    nanowires at temperatures from 75K to 125K.
    Here $G$ is the electrical conductance of the wire,
    $L_c$ is the critical length above which the lifetime may
    be approximated by $\tau \approx \nu_D^{-1} \exp(\Delta E_\infty/T)$,
    and $\Delta E_\infty$ is the activation energy for an infinitely long wire.
    Note that wires shorter than $L_c$ are predicted to have shorter lifetimes.
}
  \begin{center}
    \begin{tabular}{|c||c|c|c|c|c|}
      \hline
      \rule[-1.5ex]{0pt}{4.5ex} $G$ & $L_c $ & $\Delta E_\infty $ &
      \multicolumn{3}{c|}{$\tau$ [s]} \\
      \cline{4-6}\rule[-1.5ex]{0pt}{4.5ex}
      [$G_0$] & [\AA] & [meV] & $75\;$K & $100\;$K & $125\;$K \\
      \hline\hline
      \rule[0ex]{0pt}{3.ex}%
       3  &  2.8 & 250 & $4\times10^5$ & 2                & $5\times10^{-3}$ \\
       6  &  4.3 & 200 & 7             & $3\times10^{-3}$ & $3\times10^{-5}$ \\
       17 &  5.0 & 260 & $7\times10^5$ & 3                & $8\times10^{-3}$ \\
       23 &  6.1 & 230 & $2\times10^3$ & 0.2              & $9\times10^{-4}$ \\
       42 &  7.2 & 250 & $2\times10^5$ & 1                & $10^{-3}$ \\
       51 &  6.8 & 190 & 1             & $8\times10^{-4}$ & $10^{-4}$        \\
       67 & 18.8 & 180 & 0.6           & $5\times10^{-4}$ & $7\times10^{-6}$ \\
      \rule[-1.5ex]{0pt}{3.ex}%
       96 & 11.4 & 250 & $10^5$        & 0.8              & $3\times10^{-3}$ \\
      \hline
    \end{tabular}
  \end{center}
  \label{tab:lifetime}
\end{table}

An important prediction given in Table \ref{tab:lifetime} is that the
lifetimes of the most stable nanowires, while they do exhibit significant
variations from one conductance plateau to another, do not vary systematically
as a function of radius; the activation barriers in Table \ref{tab:lifetime}
vary by only about 30\% from one plateau to another, and the wire with a conductance of
$96 G_0$ has essentially the same lifetime as that with a conductance of
$3 G_0$.  In this sense, the activation barrier is found to be {\em universal}:  in any conductance interval, there are very short-lived wires
(not shown in Table \ref{tab:lifetime}) with very small activation barriers, while the longest-lived wires have activation barriers of a universal size
\begin{equation}\label{eq:Delta_E}
  \Delta E_\infty\, \simeq \, 0.6
    \left(\frac{\hbar^2 \sigma_s}{m_e}\right)^{1/2}\!\!\!\!\!,
\end{equation}
depending only on the surface tension of the material.
Here $m_e$ is the conduction-band effective mass, which is comparable to the free-electron rest mass.
A comparison of the lifetimes of sodium and gold nanowires \cite{Burki04b}
indicates that gold nanowires are much more stable, as expected from the larger value of the surface tension $\sigma_s({\rm Au})= 5.9\, \sigma_s({\rm Na})$.
This is consistent with the observation that gold nanowires in particular, and noble metal nanowires in general, are much more stable than alkali metal nanowires.

The fact that the typical activation energy (\ref{eq:Delta_E}) is independent of $R_0$ may be understood as follows:
The instanton is a stationary state of Eq.\ (\ref{eq:energy}); as such, the Virial theorem implies that the bending energy $\langle \frac{\kappa}{2} (\partial_z \phi)^2\rangle$ is proportional to $\langle V(\phi)\rangle$.
Since $\kappa \sim \sigma_s R_0$ and $V \sim 1/R_0$, this implies that the characteristic size of the instanton $L_c \sim \sqrt{\sigma_s} R_0$ and $\Delta E_\infty \sim \sqrt{\sigma_s}$.

The lifetimes tabulated for sodium nanowires in Table \ref{tab:lifetime}
exhibit a rapid decrease in the temperature interval between 75K and 125K.
This behavior can explain the observed temperature dependence of conductance
histograms for sodium nanowires \cite{Yanson99,Yanson00,Yanson01a},
which show clear peaks at conductances near the predicted values at
temperatures below 100K, but were not reported at higher temperatures.

\section{Conclusions}
\label{sec:conclusions}

The NFEM is the simplest possible model of metal nanostructures. Nonetheless, it is a remarkably rich model, which provides a unified description of quantum transport, stability, and structural dynamics of simple metal nanowires.
It is hoped that the {\em generic properties} of metal nanostructures elucidated by the NFEM can guide the exploration of more elaborate, material-specific models,
in the same way that the free-electron model provides an important theoretical reference point from which to understand the complex properties of real bulk metals.

\section*{Acknowledgments}
This work was supported by NSF Grant No.\ 0312028.  We thank Raymond Goldstein, Hermann Grabert, Frank Kassubek, Daniel Stein, Daniel Urban, and Chang-hua Zhang for their contributions to the work reviewed in this article.

%
\bibliographystyle{spphys}
\bibliography{refs}

\end{document}